\documentclass[traditabstract]{aa-jinst}

\usepackage{amsmath}
\usepackage{array}
\usepackage{./fguill}
\usepackage{graphicx}
\usepackage{txfonts}
\usepackage{natbib}
\bibpunct{(}{)}{;}{a}{}{,}

\newcommand\T{\rule{0pt}{2.6ex}}
\newcommand\B{\rule[-1.2ex]{0pt}{0pt}}

\begin{document}

\title{LFI Radiometric Chain Assembly (RCA) data handling ``Rachel''}
\titlerunning{~``Rachel'' -- LFI RCA data handling}

\author{M.~Malaspina\inst{1}\and E.~Franceschi\inst{1}
  \and P.~Battaglia\inst{2}\and P.~Binko\inst{4}\and R.C.~Butler\inst{1}
  \and O.~D'Arcangelo\inst{5}\and S.~Fogliani\inst{6}\and M.~Frailis\inst{6}
  \and C.~Franceschet\inst{2}\and S.~Galeotta\inst{6}\and F.~Gasparo\inst{6}
  \and A.~Gregorio\inst{7}\and M.~Lapolla\inst{2}\and R.~Leonardi\inst{8}
  \and G.~Maggio\inst{6}\and N.~Mandolesi\inst{1}\and P.~Manzato\inst{6}
  \and M.~Maris\inst{6}\and M.~Meharga\inst{4}\and P.~Meinhold\inst{8}
  \and N.~Morisset\inst{4}\and F.~Pasian\inst{6}\and F.~Perrotta\inst{6}
  \and R.~Rohlfs\inst{4}\and M.~Sandri\inst{1}\and M.~Tomasi\inst{3}
  \and M.~T\"{u}rler\inst{4}\and A.~Zacchei\inst{6}\and A.~Zonca\inst{3}
}
\authorrunning{M.~Malaspina, E.~Franceschi, et al.}

\institute{
  Istituto di Astrofisica Spaziale e Fisica Cosmica, INAF, via P.~Gobetti 101 -- I-40129~Bologna, Italy\\
  \email{malaspina@iasfbo.inaf.it}
\and
  Thales Alenia Space Italia S.p.A., IUEL - Scientific Instruments, S.S. Padana Superiore 290 -- I-20090~Vimodrone (MI), Italy
\and
  Universit\`{a} degli Studi di Milano, via Celoria 16 -- I-20133~Milano, Italy
\and
  ISDC Data Centre for Astrophysics, University of Geneva, ch. d'\'{E}cogia 16, CH-1290~Versoix, Switzerland
\and
  IFP-CNR, via Cozzi 53 -- I-20013~Milano, Italy
\and
  INAF~/~OATS, via Tiepolo 11 -- I-34143~Trieste, Italy
\and
  University of Trieste, Department of Physics, via Valerio 2 -- I-34127~Trieste, Italy
\and
  Department of Physics, University of California, Santa Barbara, CA~93106-9530, USA
}

\date{Received: June 24, 2009 / Accepted: August 7, 2009}

\abstract{\textit{Planck's Low Frequency Instrument is an array of 22 pseudo-correlation 
radiometers at 30, 44, and 70 GHz. Before integrating the overall array 
assembly, a first set of tests has been performed for each radiometer 
chain assembly (RCA), consisting of two radiometers. 
In this paper, we describe Rachel, a software application which has been 
purposely developed and used during the RCA test campaign to carry out 
both near-realtime on-line data analysis and data storage (in FITS 
format) of the raw output from the radiometric chains.
}}


\maketitle
%

\section{
Introduction}

Planck is an ESA mission designed to map the angular distribution of 
the cosmic microwave background (CMB). It will scan the sky in 9 
frequency channels, from 30 GHz to 857 GHz, with two instruments: 
the Low Frequency Instrument (LFI, 30-70 GHz) and the High Frequency 
Instrument (HFI, 100-857 GHz).\\
Planck/LFI is an array of 22 pseudo-correlation differential radiometers, 
cryogenically cooled at 20K, coupled to the telescope by 11 dual profiled 
corrugated feed horns \citep{AA_Mandolesi_M1,AA_Tauber}. Accordingly, 
Planck/LFI can be described as being composed of 11 independent units, or 
Radiometer Chain Assemblies (RCAs), each consisting of 2 pseudo-correlation 
radiometers and their 4 detectors, that is, 2 channels for each radiometer, 
switching between a sky-load and a reference-load to perform differential 
measurements \citep{AA_Bersanelli_M2}.\\
The functional and calibration test of single RCAs has been one of the 
first steps of Planck/LFI overall calibration campaign 
\citep[see][]{2005EUMA.1.3,2006SPIE.6265E..12M}. 
The main purpose of the test was to verify and measure the individual 
performance of the 11 RCAs. Two dedicated cryofacilities have been used during 
the test campaign: the 30 and 44 GHz radiometer chains were tested at Alcatel -- Alenia Space in Italy, while the 70 GHz channels were calibrated at Elektrobit in Finland.\\
From the point of view of on-line data analysis and storage, testing 
one radiometric chain implies both the acquisition and the recording, 
in form of FITS tables, of several flows of information: output 
from the four switching channels (i.e. two sky/reference pairs, 
one for each receiver), temperature probes, facility's 
setup and bias-related parameters, and user-provided log-entries. 
Moreover, a real-time graphic and numerical analysis is required, 
in order to check that the functional and environmental conditions 
of the test-in-progress meet the test requirements.\\
To perform these operations at the required rate (each channel being 
sampled at 14bits and 8192Hz, the incoming data rate was around 
0.5Mb/s), dedicated software, Rachel (Radiometric Chains Evaluator), 
has been designed and developed in an open source environment 
(Linux), using only open source libraries and tools: GCC \citep{GCC_website}, 
Qt3 \citep{QT_website}, CFITSIO \citep{CFITSIO_website}, FFTW \citep{FFTW05,FFTW_website}, 
and MySQL \citep{MySQL_website}. Rachel is open source as well, and its source 
code is available to the community, upon request (zacchei@oats.inaf.it), on the 
CVS server of the Planck/LFI Data Processing Center in Trieste.\\
Because of its high level of customizability 
and of its user-friendly interfaces, besides this Planck/LFI-specific 
application, Rachel could also be employed in many other contexts 
where data analysis and storage of one or more 180\ensuremath{^\circ} 
phase-shifted signal/reference pairs is required. \\
In this paper, the architecture and performance of Rachel are 
illustrated. Section \ref{rcaConfig} describes the test configuration.
Section \ref{rachelMain} briefly reports the software performance and architecture. 
Section \ref{realtimeIFs} illustrates the main interfaces and analysis tools 
provided by Rachel. The WEB-based remote-monitoring tool is depicted 
in Sect. \ref{rachelWeb}, while Sect. \ref{FITSfiles} describes the structure of the 
FITS tables where test data are stored.

\section{
RCA data acquisition configuration}
\label{rcaConfig}

\begin{figure}
  \resizebox{\hsize}{!}{\includegraphics{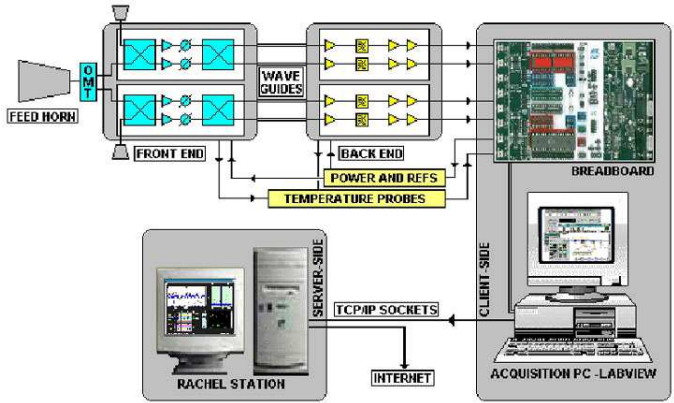}}
  \caption{Block diagram of a RCA test.}
  \label{fig1}
\end{figure}

The units involved in each RCA test (Fig. \ref{fig1}) were:
\begin{enumerate}\itemsep 0pt
\renewcommand{\labelenumi}{\arabic{enumi}.}
\item
inside the cryofacility, the RCA itself, consisting of the sky feed horn 
and a couple of small pyramidal horns, all feeding the 20K cooled Front 
End Module (FEM), and then a set of waveguides propagating signal to the 
300K Back End Modules (BEM). The feed horn and the two small antennas look, 
respectively, at an external wide band blackbody (or sky simulator) and 
at two reference loads, whose 4K temperature is provided by a dedicated 
helium cryocooler \citep{2007NewAR..51..305T};
\item
outside the cryofacility, the Data Acquisition Electronics 
(DAE) and a PC running a LabView application for raw data acquisition 
and board control;
\item
the Rachel workstation.
\end{enumerate}
In terms of information flow, transmission from FEM to BEM is via waveguide; 
from BEM to the DAE, using analog signals over wire; from DAE to the LabView 
PC, using digital signals; finally, 
from the LabView PC to the Rachel workstation, through TCP/IP 
network sockets in a client-server configuration, with Rachel 
acting as a server and LabView as a client.\\
Besides scientific data (i.e. the output from the four RCA detectors), 
network sockets are used also to acquire the output from temperature 
or current probes, the DAE configuration, and log-entries from 
the LabView PC.

\section{
Rachel's main tasks and performance}
\label{rachelMain}

Since LFI science and housekeeping data are sampled at fixed rate, 
the expected maximum average incoming data rate can be easily predicted: 
533568 b/s (see Table \ref{table:netsocks}; configuration and log-entries 
packets contribution can be safely assumed as insignificant, since 
they are sporadically generated only after manual operations). In order 
to have a margin of safety, Rachel has been designed to keep up with an 
incoming data rate up to 1 Mb/s without losing any packets. This requirement 
has been tested using a simple RCA simulator, 
``RcaClient'', expressly developed to 
generate RCA-like network packets at a user-configurable data rate. Even 
the slowest computer we used for the test, a P3@800 MHz notebook, has been 
found to be able to meet the requirement. On a P4@2.8 GHz, that is a 
machine comparable to those actually used during the RCA test campaign, 
Rachel process average CPU load with data recording enabled and 1 Mb/s 
incoming rate, measured with the Linux ``top'' utility, is between 15\% and 20\%.\\
The minimal set of tasks that Rachel is required to perform in real-time, 
while keeping up with the incoming network packets, are:
\begin{enumerate}\itemsep 0pt
\renewcommand{\labelenumi}{\arabic{enumi}.}
\item
Listening and acquiring data on a loss-less protocol (TCP/IP) 
from four network sockets, as illustrated in Table \ref{table:netsocks}.
\item
Recording incoming data as FITS tables on the local hard disk 
(see Sect. \ref{FITSfiles}).
\item
Providing graphical and numerical real-time analysis and quick-looks 
of the incoming data.
\end{enumerate}
The most CPU-consuming operation, thus the most potentially critical 
point of the whole application in terms of performance, is the production 
in near-real-time of two Hi-Freq FFT (one for sky samples, the other 
for the reference samples; see Sect. \ref{FFThigh}) and three Lo-Freq 
FFT (sky samples, reference samples, and their difference; see 
Sect. \ref{FFTlow}) per second for each channel: thus, 20 FFT/s, 
each one over a window of 4096 points. Quantitatively, using a sample 
of 110 seconds on a P4@2.8 GHz acquiring data at nominal rate (533568 b/s), 
we have profiled an average time of $1.0\pm0.1$ ms for the Hi-Freq FFTs 
and $1.0\pm0.2$ ms for the Lo-Freq FFTs: that is, 20 ms per second of 
acquisition. During the same test, CPU time for the statistical 
estimators (see Sect. \ref{statEstim}) has been found to be 0.7 ms per second, 
while the correlation calculations (see Sect. \ref{correlations}) required 
8 ms per second.

\begin{figure}[b]
  \resizebox{\hsize}{!}{\includegraphics{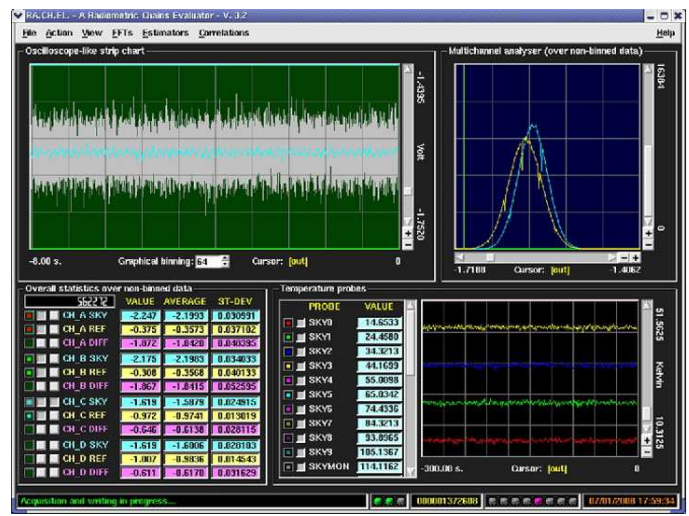}}
  \caption{Main window.}
  \label{fig2}
\end{figure}

\section{
The real-time graphic and numerical analysis interfaces}
\label{realtimeIFs}

As regards data analysis, Rachel's main task is to provide a 
quick-look of the incoming signal and of its noise properties. 
In this section, we shall illustrate the statistical tools and 
graphic interfaces which have been implemented in Rachel to achieve 
this purpose.

\begin{table*}
\caption{Network sockets}
\label{table:netsocks}
\begin{center}
\begin{small}
\begin{tabular}{|c|p{44mm}|p{85mm}|p{23mm}|}
\hline
\multicolumn{1}{|c|}{\T\textbf{Port\B}} &
\multicolumn{1}{c|}{\T\textbf{Content\B}} &
\multicolumn{1}{c|}{\T\textbf{Packet structure\B}} &
\multicolumn{1}{c|}{\parbox[t]{23mm}{\centering\T\textbf{Packet~ \\
 ~and~Bit~Rate}\B}}\\
\hline\hline
\T{18000} &
\T{Scientific data from the 4 channels.} &
\T{Synchronous network packets, each containing one packet 
counter (4 bytes), the time code (4 bytes), and 64x4 scientific 
values (unsigned 16bits words), for a total packet size of 520 bytes.}\B &
\parbox[t]{23mm}{\raggedleft\T 128 pkt/s\\[3pt]
532480 b/s\B}\\
\hline
\T{18001} &
\T{Temperature and current data from the probes.} &
\T{Synchronous network packets, each containing one 
packet counter (4 bytes), the time code (4 bytes), 
28 temperatures (4 bytes float), and 4 currents (4 
bytes float), for a total packet size of 136 bytes.}\B &
\parbox[t]{23mm}{\raggedleft\T 1 pkt/s\\[3pt]
1088 b/s\B}\\
\hline
\T{18002} &
\T{DAE configuration parameters.} &
\T{Asynchronous network packets, each containing one 
packet counter (4 bytes), the time code (4 bytes), 6 
environmental data (4 bytes float), and 32 bias data 
(1 byte), for a total packet size of 64 bytes.}\B &
\parbox[t]{23mm}{\raggedleft\T N/A\\[3pt]
N/A\B}\\
\hline
\T{18003} &
\T{Log-entries.} &
\T{Asynchronous network packets, each containing one 
packet counter (4 bytes), the time code (4 bytes), 
and one entry line (248 bytes), for a total packet 
size of 256 bytes.}\B &
\parbox[t]{23mm}{\raggedleft\T N/A\\[3pt]
N/A\B}\\
\hline\hline
\multicolumn{3}{|l|}%
{\T\textbf{Minimum average incoming data rate}\B} & 
\multicolumn{1}{r|}{\T\textbf{533568 b/s}\B}\\
\hline
\end{tabular}
\end{small}
\end{center}
\end{table*}

\subsection{
Oscilloscope-like strip chart}
\label{oscilloscope}

The oscilloscope window, located in the upper-left corner of Rachel 
(Fig. \ref{fig2}), shows a real-time plotting of channel values (Y-axis) 
over time (X-axis).\\
Each channel is represented by a user-selectable color. In order to 
disentangle overlapping channels, they can be individually obscured.
Upper and lower limits of the visible window are reported 
on the right, and can be increased or decreased by the 
{\guillemotleft}plus{\guillemotright} and {\guillemotleft}minus{\guillemotright} 
onscreen buttons (located under the scrollbar), thus allowing users to zoom 
on the signal, while the scroll-bar allows to move the window 
along the full scale range.\\
Due to the high sampling rate (8192 Hz) and to the phase-shifting 
strategy, a raw representation of the incoming data flow would 
hardly be decipherable. To improve readability, channels are 
first split into their \textit{sky}/\textit{reference} components, 
then graphically binned over a user-selectable number of 
samples, $N_{\mathrm{smp}}$, varying in the range from 4 to 512.
Thus, the X--axis extension in seconds, $\Delta X$, whose value 
is displayed in the lower left corner, is dependent on $N_{\mathrm{smp}}$
and the sampling rate $r_{\mathrm{smp}}$ (in Hz):

    \begin{equation}\label{eq:xrange}
        \Delta X = \frac{1024 N_{\mathrm{smp}} }{r_{\mathrm{smp}}}.
    \end{equation}

\noindent
Finally, whenever the cursor is located over the chart, the value 
of the pixel it is pointing at will be shown next to the \textbf{Cursor} 
label, in volt or in Kelvin according to the unit option.

\subsection{
Multi-channel analyzer distribution chart}
\label{multichannel}

The multi-channel sub-window, located on the upper-right corner of 
Rachel (Fig. \ref{fig2}), offers a graphic representation of the 
data frequency distribution. In the case of a noise-dominated signal, 
it thus allows a qualitative estimate about its Gaussianity (for a 
quantitative approach based on skewness and kurtosis, see 
Sect. \ref{statEstim}). The visible window is zoomable and scrollable 
along both the X-axis (distribution classes, or bins) and the Y axis 
(frequency density). The size of the bins can be increased or decreased 
clicking on the {\guillemotleft}plus{\guillemotright} and 
{\guillemotleft}minus{\guillemotright} buttons next to the horizontal 
scrollbar. As regards bin size, the lower limit corresponds to the LSB 
of the ADCs (one bin per pixel), while the upper limit is set to make 
the maximum ADC range coincide with the window width.\\
In order to keep the distribution curve in range, the Y-axis scale is 
automatically increased whenever the frequency density of the mode 
reaches the upper limit. For the same purpose, an auto-zooming algorithm 
has been implemented which adjusts automatically the vertical scale 
whenever the size of the bins is decreased. Obviously, both these 
mechanisms can be overridden by users clicking the 
{\guillemotleft}plus{\guillemotright} and {\guillemotleft}minus{\guillemotright} 
buttons next to the vertical scrollbar: this can be very useful when 
the four channels have show different behaviors, or to deal with spikes 
in the frequency density.\\ 
Finally, as in the oscilloscope window, each channel is represented by 
a user-selectable color, and a \textbf{Cursor} label will show, numerically, 
the frequency density of the bin pointed by the cursor. Hence, it is 
possible to have, at a glance, a quick estimate of the statistical 
dispersion of the samples.

\subsection{
Strip chart of the temperature probes}
\label{stripChart}

The strip chart of the temperature probes allows the user to keep track 
of any variation in the environmental condition during the latest five 
minutes of acquisition. The chart itself is functionally similar to the 
oscilloscope tool (see Sect. \ref{oscilloscope}), and the latest value 
of each probe is numerically reported as well.\\
Actually, the choice of an unchangeable five minutes interval has been 
rather arbitrary: initially, it had been dictated merely by the width 
of the window (300 pixels) and the temperature probes sampling rate 
(1 Hz), thus resulting in 300 seconds of non-binned data. Anyway, 
since during the first tests an interval of five minutes has proved 
satisfactory for checking temperature trend and stability, we have 
never implemented a feature to make it user changeable.

\subsection{
Overall statistics}
\label{overallStats}

The overall statistics sub-window, located on the lower-left 
corner of Rachel (Fig. \ref{fig2}), displays numerical information 
about the scientific channels.\\
Each channel is split into its \textit{sky} 
and \textit{reference} components, and the difference between them 
is reported as well. The \textbf{VALUE} columns show the latest value, 
while the \textbf{AVERAGE} and the \textbf{ST-DEV} columns show the average 
and the biased standard deviation of \textit{non-binned samples} over 
all the data acquired since \textbf{Reset statistics} was last invoked.
The LCD-like figures in the upper-left corner of the statistics sub-window 
is the sample counter, and it is zeroed every time that statistics are reset.
Statistics are directly performed on the incoming 
data flow, and are refreshed every 0.5s.\\
The check boxes on the left of the channel names affect the behavior 
of the oscilloscope: they allow the user to enable or disable 
the plotting of the selected channel component and of its dispersion 
inside the graphical bin.

\subsection{
High frequency FFT}
\label{FFThigh}

\begin{figure}[h]
  \sidecaption
  \resizebox{\hsize}{!}{\includegraphics{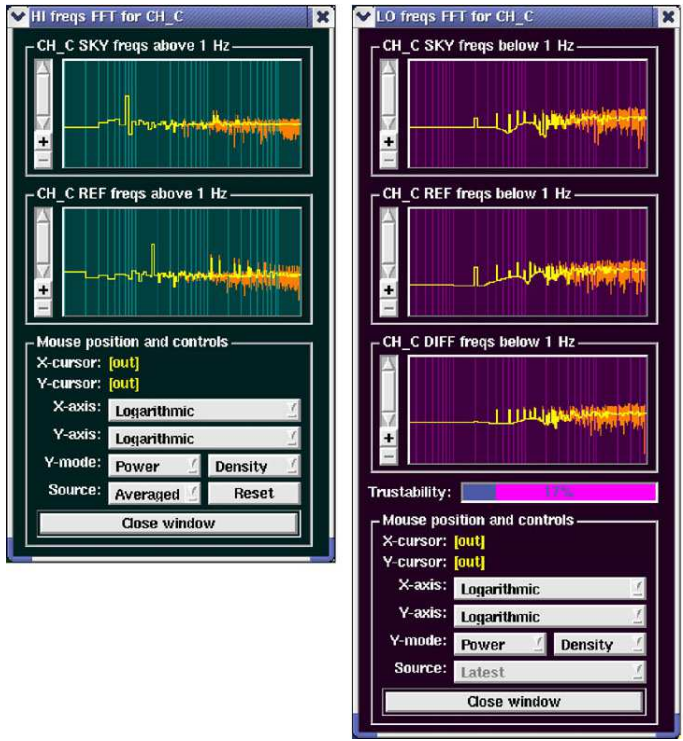}}
  \caption{High Freq FFT (left) and Low Freq FFT (right) dialog boxes.}
  \label{fig3}
\end{figure}

Each high frequency FFT dialog box (Fig. \ref{fig3}, on the left) 
displays the power spectrum, for components above 1Hz, of the \textit{sky} 
and \textit{reference} values of one scientific channel. Since one RCA has 
four channels, as explained in the Introduction, Rachel provides 
four independent Hi-Freq FFT windows. Users can toggle the visibility 
of each of these windows through the FFT menu.\\

\noindent
The FFTs are performed over the latest 8192 samples/channel. \textit{Sky} 
and \textit{reference} values being treated separately, each FFT takes 
as input 4096 homogeneous samples, that is one second of data, 
and produces a vector of components which ranges from 1 to 
2047 Hz (\ensuremath{\frac12 f}$_{Nyquist}$ -1).\\
Since in the RCA tests frequencies above 1Hz are expected to 
show a flat spectrum, these dialogs can be useful to detect unexpected 
components---notably, the 50 Hz interference from power supplies 
and microphonic noise. Users can choose whether to represent 
frequencies over a logarithmic or a linear axis, while amplitudes 
can be shown both in Volts or in dBV:

\vspace{-1mm}
    \begin{equation}\label{eq:Vampl}
        A_{\mathrm{V}} = \frac{1}{N} \sqrt{R_{\mathrm{FFT}}^2 + I_{\mathrm{FFT}}^2}
    \end{equation}
\vspace{-1mm}
    \begin{equation}\label{eq:dBV}
        A_{\mathrm{dBV}} = 20 \log \left( A_{\mathrm{V}} \right) \hspace{7mm}
    \end{equation}

\vspace{2mm}
where $R_{\mathrm{FFT}}$ and $I_{\mathrm{FFT}}$ are, respectively, the real and 
the imaginary part of the FFT output, and \textit{N} is the number of components.\\
On the left of each graph there is a zooming bar which allows 
the user to increase or decrease the Y-axis range. Whenever the 
cursor is located over one of the two charts, the frequency and 
the amplitude of the pixel it is pointing at will be shown next 
to the \textbf{X-cursor} and \textbf{Y-cursor} labels, thus allowing the user 
to single out frequency classes which show anomalous peaks.

\subsection{
Low frequency FFT}
\label{FFTlow}

Low frequency FFT dialog boxes (Fig. \ref{fig3}, on the right) are similar 
to High Frequency FFTs, the main difference being that they take into 
account the 0.2-500 mHz components. Furthermore, besides the \textit{sky} 
and \textit{reference} windows, they provide a third FFT window for the 
\textit{difference} between \textit{sky} and \textit{reference}. Their 
main purpose is to provide a rough outline of the 1/\textit{f} noise 
contribution, the 1/\textit{f} knee frequency expected for the LFI 
detectors being below 50 mHz \citep{1999A&AS..140..383M}.

\begin{figure}[t]
  \resizebox{\hsize}{!}{\includegraphics{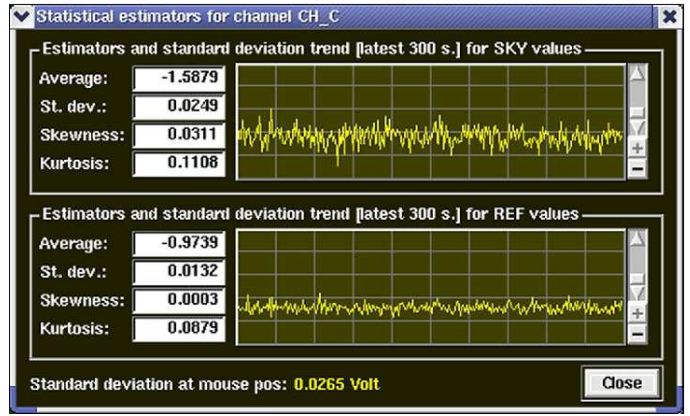}}
  \caption{Statistical estimators dialog box.}
  \label{fig4}
\end{figure}
\begin{figure}[b]
  \includegraphics[width=80mm]{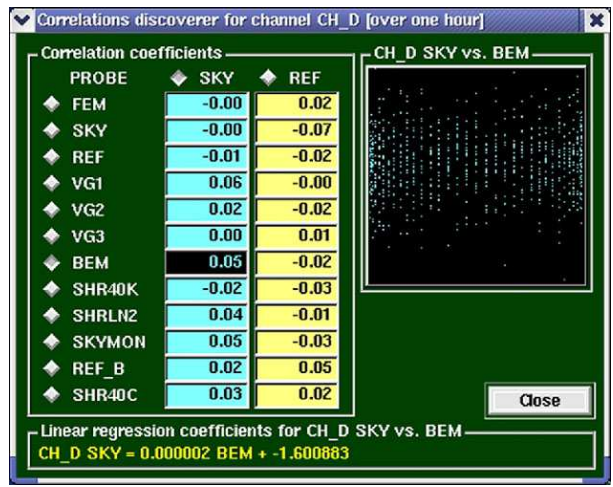}
  \caption{Correlations discoverer dialog box.}
  \label{fig5}
\end{figure}

\subsection{
Statistical estimators}
\label{statEstim}

The statistical estimator dialog boxes (Fig. \ref{fig4}) provide, for the 
latest five minutes of acquisition, real-time statistical estimators 
(refresh frequency: 1Hz) for each channel's \textit{sky} and \textit{reference} 
pairs: average, standard deviation, skewness, and kurtosis.\\
Skewness and kurtosis, giving a numerical estimate of the degree of the 
asymmetry and of the peakedness of the distribution, allow the user to quantify 
the (non-)gaussianity of the incoming signals. The graph of the standard 
deviation trend can be useful to detect noise instabilities and fluctuations. 
Five minutes is a convenient interval to graphically compare noise fluctuations 
with the values displayed in the temperature strip charts (Sect. \ref{stripChart}).

\subsection{
Correlations discoverer}
\label{correlations}

The correlations discoverer dialog boxes (Fig. \ref{fig5}) calculate 
and represent in real-time (refresh frequency: 1Hz) the matrix of correlation 
coefficients between incoming scientific data and temperature 
probes. They also calculate the linear regression, and display 
a correlation graph for the selected channel/probe pair.\\
This tool is extremely useful to detect at a glance both unexpected 
correlations (due, for instance, to unforeseen thermal couplings) 
and the absence of predictable correlations. 

\begin{figure}[b]
  \resizebox{\hsize}{!}{\includegraphics{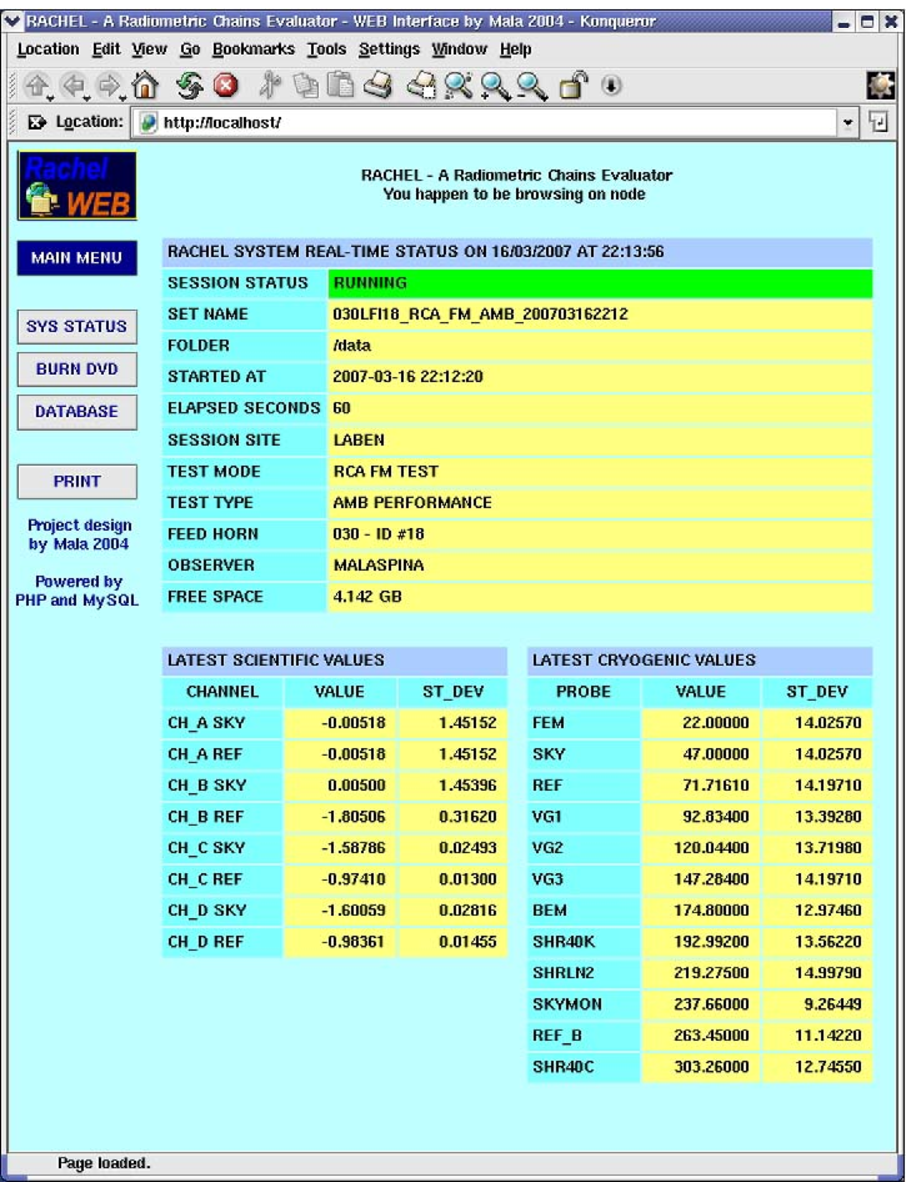}}
  \caption{WEB-based remote interface.}
  \label{fig6}
\end{figure}

\section{
SQL-based WEB interface for remote monitoring}
\label{rachelWeb}

Because of the long duration of many RCA tests and the consequent need to
monitor the behavior of the chains also from remote facilities, a SQL-based
WEB interface has been added to Rachel (Fig. \ref{fig6}).\\
An interface, known as ``Rachel-WEB'', displays a numerical quick-look 
of the on-going test and its
main parameters (science data, currents, and temperatures). The refresh
frequency of Rachel-WEB data is 1Hz, thus it acts as a simplified
near-realtime quick-look.\\
In order to minimize the impact of this interface on Rachel's 
performance, averaged data are stored into a MySQL database 
through low priority buffered queries, which do not 
introduce wait states. The same database has 
also been used to keep track of the trend of each test, thus 
providing a permanent, shared and searchable catalogue to 
retrieve FITS files from one or more sessions of test.

\section{
FITS files production}
\label{FITSfiles}

\begin{figure}[t]
  \resizebox{\hsize}{!}{\includegraphics{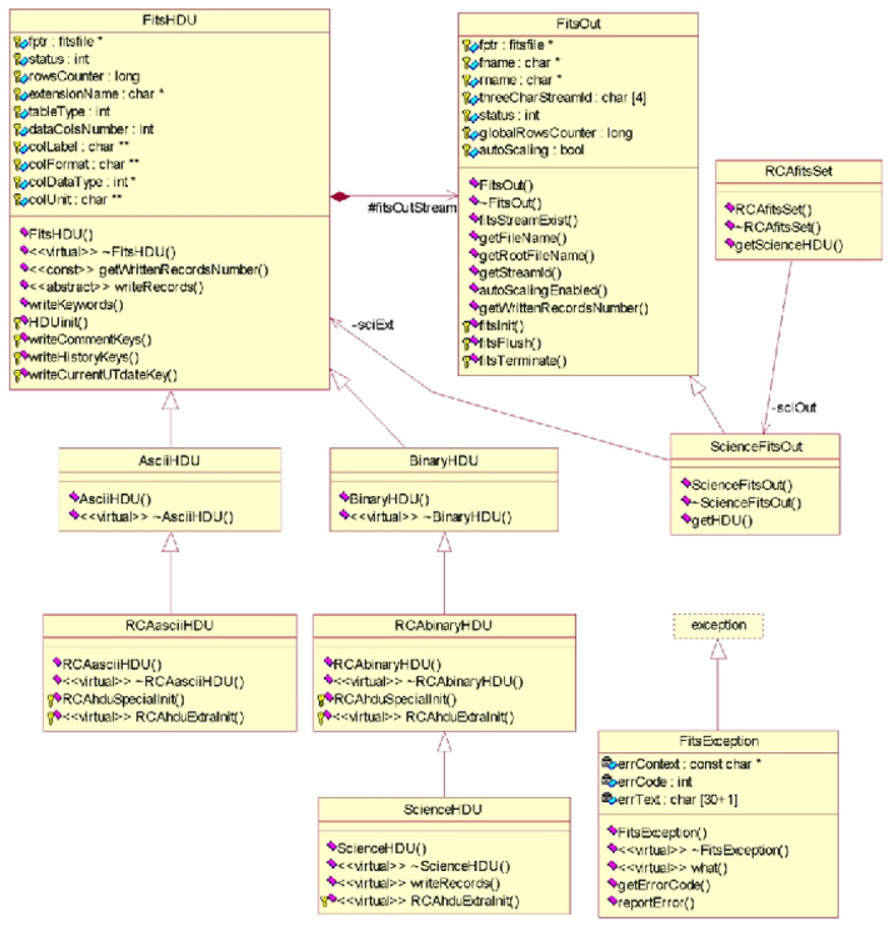}}
  \caption{UML sample of the \textit{fitswriter} classes.}
  \label{fig7}
\end{figure}

In order to make test results available to the off-line analysis tools 
and DB \citep{PL-LFI-OAT-AD-003}, data from RCAs calibration tests are 
stored in FITS format.\\
As specified by LFI DPC data management requirements, four separate 
files are generated for each test: each file contains a collapsed primary 
array and one Header and Data Unit \citep{PL-LFI-OAT-AD-004}. The Data Unit 
contains data from one network socket only, while the Header contains a 
set of \textit{card images} (or \textit{Keyword Records}, as they have been
renamed in the latest revision of the FITS standard) which is shared by all 
the files.\\
The module which provides Rachel on-line FITS storage functionality is 
called \textit{fitswriter}. Basically, fitswriter is a C++ wrapper built 
around the CFITSIO library \citep[see][]{CFITSIO_website}.\\
Its goal is to maintain the maximum control over the implementation 
details and to meet the FITS format specifications given above. 
Moreover, through a clear-cut separation between the project-dependent 
classes and the generic ones, it guarantees a high reusability of the code.\\
A graphical sample of the fitswriter classes, expressed in 
Unified Modeling Language, is given in Fig. \ref{fig7}.

\section{
Conclusions}
\label{Conclusions}

All the data acquired during the RCAs tests carried out at Alcatel Alenia 
Space (Italy) during 2004 and at Elektrobit (Finland) in the summer of 2005 
have been successfully processed and viewed using Rachel.\\
Many issues in the test data have been detected and efficiently managed 
thanks to its advanced quick-look interface.\\
About 220 gigabytes of FITS files have been produced and made available for 
the subsequent offline data analysis phase \citep[see][]{JI_Tomasi}.

\vspace{4mm}
\begin{acknowledgements}
Planck is a project of the European Space Agency with instruments
funded by ESA member states, and with special contributions from Denmark
and NASA (USA). The Planck-LFI project is developed by an International
Consortium lead by Italy and involving Canada, Finland, Germany, Norway,
Spain, Switzerland, UK, USA.\\
The Italian contribution to Planck is supported by the Italian Space Agency (ASI).
The US Planck Project is supported by the NASA Science Mission Directorate.
\end{acknowledgements}

\bibliographystyle{aa-jinst}  
\bibliography{JINST_013T_0609_astro-ph} 

\end{document}